\documentstyle[12pt]{article}
\begin{document}
\begin{center}
to appear in Physical Review Letters
\end{center}
\begin{flushright}
IITAP-96-12-01 \\
December 1996  \\
Extended version: March 1997 \\
hep-th/9612147
\end{flushright}
\vspace*{2.5cm}
\begin{center}
{\large Effective Regge QCD} \\
\vspace*{2.5cm}
{\large  Victor T. Kim${}^{\S \dagger}$
and Grigorii B. Pivovarov${}^{\S \ddagger}$ }\\
\vspace*{0.3cm}
{\em $\S $ : International Institute of Theoretical and Applied Physics,\\
     Iowa State University, Ames, Iowa 50011-3022, USA }\\
\vspace*{0.3cm}
{\em ${}^\dagger$ :
St.Petersburg Nuclear Physics Institute,
 188350 Gatchina,
Russia} \footnote{permanent address; e-mail: $kim@pnpi.spb.ru$}\\
\vspace*{0.3cm}
{\em ${}^\ddagger$ :
Institute for Nuclear
Research, 117312 Moscow, Russia}
\footnote{ permanent address; e-mail: $gbpivo@ms2.inr.ac.ru$}
\end{center}
\vspace{1cm}
\begin{center}
{\large \bf Abstract}
\end{center}

A new framework for a high energy limit of quantum 
gauge field theories is introduced.
Its potency is illustrated on a new derivation
of the reggeization of the gluon.

\vspace*{1.5cm}
PACS number(s): 11.10.Hi, 11.10.Ji, 11.55.Jy, 12.38.Cy

\newpage

The concept of effective field theory (for a review, see \cite{Pol92})
is an important means to obtain qualitative understanding of   
quantum field theories. It is also an important 
calculational tool. 

Effective field theory provides a mechanism of 
unification---the plethora of its fields
is consolidated by this mechanism into a more 
compact set of fields of the underlying theory which implements a  
symmetry richer than the one of the effective theory. A well known
example is the connection between the standard model and the 
grand unification
models \cite{GUT}. Another example is nonrelativistic QED
 \cite{Lep86,Piv94},
where a single fermion field breaks down into two independent fields
of fermions and antifermions. 

The calculational advantage is obtained via the usage of the renormalization 
group (RG) invariance \cite{Wil74}. It   
requires the predictions of the effective theory to be independent
of the particular way the fields of the underlying theory were broken into
the fields of the effective theory. Solving the RG equations 
is the most effective way to make a resummation of 
the perturbative expansion for the underlying theory taken in the 
corresponding ``leading logarithm approximation'' (for the notion of 
the leading logarithms see, for example, \cite{LLA}).  

The necessity to develop an effective theory for Regge limit
of gauge theories is well acknowledged [7-10].
 It is more
pressing now as the energy reached in hadron collisions 
provides data (see, e.g., \cite{CDF93}) whose understanding 
requires \cite{Gie96} 
an account of the resummation of the leading energy logarithms for QCD.
For known results on the resummation see \cite{Lip96,Lip76}. 
Attempts of such an account are available \cite{Mue87}.

Here we present a new approach which is, in our opinion, the
most straightforward realization of the effective field theory
concept for the Regge limit. It allows one to prove the reggeization
of the gluon \cite{Lip76,Gri73}---a property which 
is, at the moment, a well-tested conjecture. 

Consider a Green's function of a field theory,
$$G(x_1,...,x_n)=\,<T\Phi(x_1)...\Phi(x_n)>,$$ where any $x_n$
comprises all the variables labelling the field $\Phi$ (in particular,
the space-time coordinates or the momenta if one chooses the momentum 
representation). Consider next the Green's function of the boosted
fields,
\begin{equation}
\label{bG}
G_{\lambda}(x_1,...,x_n;y_1,...,y_m)=\,<T\Phi_{\lambda}(x_1)...
                                         \Phi_{\lambda}(x_n)
                                         \Phi_{1/\lambda}(y_1)...
                                         \Phi_{1/\lambda}(y_m)>,
\end{equation}
where $\Phi_{\lambda}(x)=B^{-1}(\lambda)\Phi(B(\lambda) x)$ is the Lorentz
boost of the field $\Phi$ along a $z$-axis, parameterized by the exponential
of the rapidity of the boost,
$\lambda=\sqrt{(B(\lambda)p)_{+}/(B(\lambda)p)_{-}}$, for a four-vector
$p$ ($p_{\pm}=p_0\pm p_z$ are the light-cone components of $p$). The Regge 
limit is then
the one of infinite rapidity, i.e., $\lambda\rightarrow\infty$. The effective
Regge theory is to approximate the boosted Green's functions of 
Eq. (1) at large $\lambda$.

Intuitively, the collection of the fields labeled by $x_i$ in 
Eq. (1) represents the excitations moving fast rightwards
in the $z$-direction, while the $y$-fields represent the fast left-movers.
The $\lambda$ then scales as  the invariant energy
of the relative motion of the left- and right-movers.

The momenta of the fields in the right hand side (rhs) of Eq. (1)
satisfy, for large $\lambda$, either
\begin{equation}
(\mid p_{+}\mid>\mu_1, \,\mid p_{-}\mid<\mu_2)\Leftrightarrow (p\in\Omega_R)
\end{equation}
for the right-movers, or
\begin{equation}
(\mid p_{+}\mid <\mu_2, \,\mid p_{-}\mid>\mu_1)\Leftrightarrow (p\in\Omega_L)
\end{equation}
for the left-movers. The $\mu_i$ are  arbitrary scales. The only 
requirement we need on them is $\Omega_R\cap\Omega_L=\emptyset$.
The RG invariance
requires that any physical prediction be independent of their values.

Instead of the single field $\Phi$ of the underlying theory, 
the effective theory has two independent fields,
\begin{equation}
\Big( R_{1/\lambda}(x)=\Phi(x) \Big)_{\mid_{p\in\Omega_R}},\,
\Big( L_{\lambda}(x)=\Phi(x) \Big)_{\mid_{p\in\Omega_L}}.
\end{equation}
The $p$ above is supposed to belong to $x$. The $\lambda$-subscript
denotes the boost transformation, as in Eq. (1). 

With Eq. (4), the boosted Green's function of Eq. (1) is
\begin{equation}
G_{\lambda}(x_1,...,x_n;y_1,...,y_m)=\,<TR(x_1)...
                                         R(x_n)
                                         L(y_1)...
                                         L(y_m)>.
\end{equation}   

The effective action for the fields $R,L$ is defined
by
\begin{equation}
\exp(iS_{eff}(R,L,\lambda))=\int \prod_{p\in{\!\!\!/}(\Omega_R\cup\Omega_L)}
d\Phi(x)
\,\exp(iS(\Phi)). 
\end{equation}
Where the rhs is expressed through $R,L$ by Eq. (4). 

If the boosted Green's functions of Eq. (1) are finite in the limit 
$\lambda\rightarrow\infty$ modulo logarithms of
$\lambda$, i.e., logarithms of the invariant energy for 
the relative motion of the left- and right-movers, 
then the effective action of Eq. (6) should have a finite limit,
\begin{equation}
S_{eff}(R,L)=\lim_{\lambda\rightarrow\infty}S_{eff}(R,L,\lambda).
\end{equation}
In that case, the only dependence on $\lambda$ left in the effective 
theory enters through the scales $\mu_i$ of Eqs. (2),(3) which cut
the effective theory since after been boosted the scales get a 
$\lambda$-dependence. This removes the cuts from the effective
theory in the limit $\lambda\rightarrow\infty$. In this way, 
to study the energy dependence in the 
Regge limit means to study the infrared and ultraviolet divergences of 
the effective theory.

Note that the cuts involve only the longitudinal directions. Thus we need
to study only the divergences of a {\it 2-dimensional} (2D) field theory
at fixed values of the transverse coordinates (momenta).
By transverse coordinates we mean all the variables labelling the $R$ and
$L$ fields but the light-cone coordinates $x_{\pm}$. The coupling 
``constants'' of this  2D theory depend on the transverse coordinates.
The RG equations for the coupling constants are an integral equations
in the space of the transverse coordinates. In the case of QCD,
these equations contain the same information as the BFKL equation 
\cite{Lip76}.   

To check this, consider the gluodynamics as the underlying theory, i.e.,
\begin{equation}
S_{glue}(A)=-\frac{1}{4}(F_{\mu\nu}^a)^2,
\end{equation}
with $A$ denoting the color octet of vector gluon fields, 
$F_{\mu\nu}^a=\partial_\mu A_\nu^a-\partial_\nu A_\mu^a-
gC_{abc}A_\mu^bA_\nu^c$ and
$g$ the gauge coupling.
The effective Regge gluodynamics is then a theory of two color octets 
of vector fields $R$ and $L$. With Eq. (6), 
its action up to terms of order $g^2$ is  
\begin{equation}
S_{eff}(R,L) = S_{glue}(R) + S_{glue}(L) + S_{int}(R,L), 
\end{equation}
where $S_{int}$ is an action of interaction between the right- and
left-boosted gluons, bilinear in $R$ and $L$.

Note that the self-interaction of both right- and left-movers mimics
the self-interaction of the underlying field. This is a consequence of
the Lorentz invariance of the underlying theory.

The $S_{int}$-term comes from the diagrams in Fig. 1. Infinitely 
boosting the fields according with the labels on the external legs of the 
diagrams from Fig. 1, one obtains a finite expression for $S_{int}$. 
The finiteness
is a nontrivial outcome of a cancellation of infinities between the 
contributions of Fig. 1(b) and Fig. 1(c).

Intermediate steps of the following calculation may depend on the gauge.
In these cases Feynman gauge is implied.  

To specify the form of $S_{int}$ from Eq. (9), we need the following
objects:
\begin{equation}
({\mathcal N}^R)_{m}(x_{\perp})=\int\,dx_{-}dy_{-}
\partial_i R_{+}^a(x_-,x_{\perp})D(x_- - y_-)
\partial_i R_{+}^b(y_-,x_{\perp})C_{mba},
\end{equation}
where $R_{\mu}^a$ is the gluon right-movers 
field taken on a light-front $x_+=0$, $\partial_i$ are the derivatives
over transverse coordinates (summation over $i=1,2$ is implied), and $D$ is 
\begin{equation}
D(x)=\int\frac{dk}{2\pi}\frac{e^{ikx}}{ik};
\end{equation}
\begin{equation}
({\mathcal M}^R)_{m}(x_{\perp})=\int\,dx_{-}dy_{-}
\partial_i\partial_i
(R_{+}^a(x_-,x_{\perp})D(x_- - y_-)R_{+}^b(y_-,x_{\perp}))C_{mba},
\end{equation}
\begin{eqnarray}
({\mathcal A}_1^R)_{m}(x_{\perp})=\int\,dx_{-}
(\partial^- R_{+}^a(x_-,x_{\perp})R_{-}^b(x_-,x_{\perp})+
\partial^- R_{-}^a(x_-,x_{\perp})R_{+}^b(x_-,x_{\perp})+\nonumber\\
+R_{+}^a(x_-,x_{\perp})\partial^+R_{+}^b(x_-,x_{\perp})
-\partial^+R_{+}^a(x_-,x_{\perp})R_+^b(x_-,x_{\perp}))C_{mba},
\end{eqnarray} 
\begin{equation}
({\mathcal A}_2^R)_{m}(x_{\perp})=\int\,dx_{-}
(\partial^iR_{i}^a(x_-,x_{\perp})R_{+}^b(x_-,x_{\perp})
-R_{+}^a(x_-,x_{\perp})\partial^iR_{i}^b(x_-,x_{\perp}))C_{mba},
\end{equation} 
\begin{equation}
({\mathcal A}_3^R)_{m}(x_{\perp})=\int\,dx_{-}
(R_{i}^a(x_-,x_{\perp})\partial^iR_{+}^b(x_-,x_{\perp})-
\partial^iR_{+}^a(x_-,x_{\perp})R_{i}^b(x_-,x_{\perp}))C_{mba},
\end{equation} 
\begin{equation}
({\mathcal A}_4^R)_{m}(x_{\perp})=\int\,dx_{-}
\partial^- R_{i}^a(x_-,x_{\perp})R_{i}^b(x_-,x_{\perp})C_{mba}.
\end{equation}
The derivatives in the above equations act only on their nearest right
neighbors. For further convenience, the following Fourier transformations
of linear combinations of the above operators are intoduced:
\begin{eqnarray}
\tilde{\mathcal N}^R_m(q_\perp)=\frac{\pi}{2}
\int\frac{d^{2}x_\perp}{(2\pi)^2}e^{-iq_\perp x_\perp}
(\frac{1}{2}({\mathcal N}^R)_m(x_\perp)+
\frac{1}{4}({\mathcal M}^R)_m(x_\perp)
-\frac{1}{2}({\mathcal A}_1^R)_m(x_\perp)-\nonumber\\
-\frac{1}{2}({\mathcal A}_2^R)_m(x_\perp)
-2({\mathcal A}_3^R)_m(x_\perp)
+4({\mathcal A}_4^R)_m(x_\perp)),
\end{eqnarray}
\begin{eqnarray}
\tilde{\mathcal M}^R_m(q_\perp)=\frac{\pi}{2}
\int\frac{d^{2}x_\perp}{(2\pi)^2}e^{-iq_\perp x_\perp}
(\frac{1}{2}({\mathcal N}^R)_m(x_\perp)-
\frac{1}{4}({\mathcal M}^R)_m(x_\perp)
-\frac{1}{2}({\mathcal A}_1^R)_m(x_\perp)-\nonumber\\
-\frac{1}{2}({\mathcal A}_2^R)_m(x_\perp)
-2({\mathcal A}_3^R)_m(x_\perp)
+4({\mathcal A}_4^R)_m(x_\perp)),
\end{eqnarray}
\begin{equation}
\tilde{\mathcal J}^R_m(q_\perp)=\frac{\pi}{2}
\int\frac{d^{2}x_\perp}{(2\pi)^2}e^{-iq_\perp x_\perp}
(({\mathcal A}_1^R)_m(x_\perp)
+({\mathcal A}_2^R)_m(x_\perp)
+2({\mathcal A}_3^R)_m(x_\perp)
-4({\mathcal A}_4^R)_m(x_\perp)),
\end{equation}
\begin{equation}
\tilde{\mathcal L}^R_m(q_\perp)=\frac{\pi}{2}
\int\frac{d^{2}x_\perp}{(2\pi)^2}e^{-iq_\perp x_\perp}
(({\mathcal A}_1^R)_m(x_\perp)
+({\mathcal A}_2^R)_m(x_\perp)
+4({\mathcal A}_3^R)_m(x_\perp)
-4({\mathcal A}_4^R)_m(x_\perp)).
\end{equation}
We also need the same set of objects for the left-movers;
the definitions may be contrived from Eqs. (10)-(20) 
by the substitutions
$R\rightarrow L, +\rightarrow -, -\rightarrow+$ (the last two
substitutions act on the longitudinal Lorentz indices).

$S_{int}$ is expressed in terms of these objects as 
\begin{eqnarray}
S_{int}(R,L)=\int d^{2}q_\perp(G_0(q_\perp)
(\tilde{\mathcal J}^R_m(q_\perp)\tilde{\mathcal J}^L_m(-q_\perp)
-\tilde{\mathcal M}^R_m(q_\perp)\tilde{\mathcal M}^L_m(-q_\perp)-\nonumber\\
-\tilde{\mathcal M}^R_m(q_\perp)\tilde{\mathcal L}^L_m(-q_\perp)
-\tilde{\mathcal L}^R_m(q_\perp)\tilde{\mathcal M}^L_m(-q_\perp))+\nonumber\\
+G_1(q_\perp)
(\tilde{\mathcal N}^R_m(q_\perp)\tilde{\mathcal L}^L_m(-q_\perp)
+\tilde{\mathcal L}^R_m(q_\perp)\tilde{\mathcal N}^L_m(-q_\perp))+\nonumber\\
+G_2(q_\perp)
\tilde{\mathcal N}^R_m(q_\perp)\tilde{\mathcal N}^L_m(-q_\perp)),
\end{eqnarray}
where summation over colour indices $m$ is implied. 

The   ``coupling constants''
$G_{k=0,1,2}(q_\perp)$ depend on the transverse momentum transferred 
from the right-movers to the left-movers:
\begin{equation}
  G_k(q_\perp) = \frac{g^2}{-q_\perp^2}.
\end{equation}
We need to distinguish between them because they couple different
number of ${\mathcal N}$-operators.

The first ${\mathcal J}^R{\mathcal J}^L$-term of Eq. (21) is the contribution
of Fig. 1(a) originating from the interaction between the 
left- and right-movers via exchange of a gluon which momentum 
has $p_\pm<\mu_1$.

It is important to note the following: right-movers interact with 
the left-movers only by their values at the light-front $x_+=0$,
while the left movers interact only by their values on the perpendicular
light-front $x_-=0$. It is also important that ${\mathcal N}$, ${\mathcal M}$
 are nonlocal
on the corresponding light-fronts, while ${\mathcal A}_i$ are local.

We stress that Eq. (21) is only the leading order contribution
to the expansion over the coupling constant of the action of 
interaction between the left- and right-boosted gluons. It remains an 
important unsolved problem to establish the general form of this
action of interaction in higher orders of and beyond the perturbation theory.

Next we consider the divergences of the one-loop Feynman diagrams
generated by $S_{eff}$ of Eq. (9). We need to consider the divergences
in the longitudinal integrations arising as the cuts on the longitudinal
momenta from Eq. (2) for the $R$-field and from Eq. (3) for the $L$-field
are removed. That should be done at fixed transverse coordinates. 

It turns out that there are only two divergent diagrams in the one-loop
approximation
(Fig. 2). They are logarithmically divergent in the infrared because
of the singularity $1/ik$ in the rhs of Eq. (11). As it can be seen in the 
diagrams, the problem is factorized: calculation of the divergences
of Fig. 2(a) does not involve the $L$-fields, while calculation
of Fig. 2(b) does not involve the $R$-fields. Thus, 
it suffices to study the infrared divergences generated by 
the self-interaction of the $R$-fields in the operators 
$\tilde{\mathcal N}_m^R$, $\tilde{\mathcal M}_m^R$ of Eqs. (17),(18). 
The left sector gives the same divergences.

It is a remarkable fact that these infrared divergences may be absorbed
in a multiplicative renormalization of the 
operators $\tilde{\mathcal N}_m^{R,L}$. Namely, a simple calculation
shows that
\begin{equation}
d.p.\langle\tilde{\mathcal N}_m^{R,L}(q_\perp)\rangle=
\log (\lambda)
\alpha(q_\perp)\tilde{\mathcal N}_m^{R,L}(q_\perp),
\end{equation}
where $d.p.$ means {\it divergent part} and the angle brackets
denote the correction of the ${\mathcal N}_m^R$ (${\mathcal N}_m^L$) 
for the self-interaction of the $R$-fields ($L$-fields), and 
$\lambda$ is the exponential of the rapidity of the boost.
The rest of the operators (${\mathcal M},{\mathcal J},{\mathcal L}$)
participating in Eq. (21) are finite in the one-loop approximation.
We should note that Eq. (23) was obtained with the dimensional regularization
of the integration over transverse momentum involved (in particular,
integrals like $\int dl_\perp/l_\perp^2$ was set to zero).  

The $\alpha(q_\perp)$ of Eq. (23) determines the leading contribution
to the renormalization ``constant'' of the operators 
$\tilde{\mathcal N}_m^{R,L}(q_\perp)$. It turns out to be
\begin{equation}
\alpha(q_\perp)=3\alpha_Sq_\perp^2\int
\frac{d^{2}k_\perp}{(2\pi)^2k^2_\perp(k-q)_\perp^2}
\end{equation}
coinciding with the known Regge trajectory of the reggeized gluon
\cite{Lip76,Gri73}.

It follows from Eq. (23) and the one-loop finiteness
of the operators ${\mathcal M},{\mathcal J},{\mathcal L}$
that all one-loop divergences generated
by the effective action of Eq. (21) may be removed by
a renormalization of the coupling constants, 
$G_k(q_\perp)\rightarrow G_{k,R}(q_\perp)=Z_k(q_\perp)G_k(q_\perp)$. 
The renormalization
constants $Z_k(q_\perp)$, $k=0,1,2$, are
\begin{equation}
Z_k(q_\perp)=1-k\log(\lambda)\alpha(q_\perp)
\end{equation}
in the one-loop approximation.

This allows one to apply the standard
RG considerations (see, e.g., a text-book \cite{Ste93}). 
In particular, running couplings $\bar G_k(q_\perp)$ may be defined  as 
functions of the logarithm of the boosted 
cut $\mu_1/\lambda$. The standard
procedure yields the following RG equation for the running couplings:
\begin{equation}
\frac{\partial \bar G_k(q_\perp)}{\partial\log(\lambda)}=
-\frac{1}{Z_k(q_\perp)}\frac{\partial Z_k(q_\perp)}{\partial\log(\lambda)}
\bar G_k (q_\perp)\approx
k\alpha(q_\perp)\bar G_k (q_\perp),
\end{equation}
which solution is 
$$\bar G_k (q_\perp)=\lambda^{k\alpha(q_\perp)}\frac{g^2}{-q_\perp^2}$$
(this takes into account that boundary values of $\bar G_k (q_\perp)$
at $\lambda=1$ are given by Eq. (22)).

To conclude, we formulated an effective Regge gauge field theory
and recognized the problem of resummation of energy logarithms 
for gauge theories as a problem of infrared renormalization
of some nonlocal operators in a 2D field theory. The trajectory
of the reggeized gluon was rederived in this way. In a forthcoming
article \cite{KP97} we shall show that the same renormalization
of the nonlocal operators contains, in the two-loop approximation,
the BFKL equation.

 We thank I.Ya. Aref'eva, I.I. Balitsky, B.I. Ermolaev, 
 L.D. Faddeev, I.F. Ginzburg, V.A. Kuzmin, E.M. Levin, L.N. Lipatov,
 V.A. Matveev, N.N. Nikolaev, A. Petridis, J.P. Vary, and A.A. Vorobyov 
 for stimulating discussions.
 We are grateful to the Fermilab Theory Division for warm hospitality.
 V.T.K. is  thankful to the Aspen Center for Physics
 for its hospitality. G.B.P. wishes to thank L. McLerran 
 and R. Venugopalan for fruitful discussions,
 and also the Institute for the Nuclear Theory of
 the  University of Washington, Seattle and the Nordita, Copenhagen
 for their kind hospitality.
 This work was supported in part by the Russian Foundation
 for Basic Research, grants No. 96-02-16717 and 96-02-18897.

\vspace*{2cm}
\section*{Figure Captions}
Fig. 1. Diagrams contributing to $S_{int}$ to order $g^2$
\newline
Fig. 2. Divergent one-loop diagrams of the effective theory

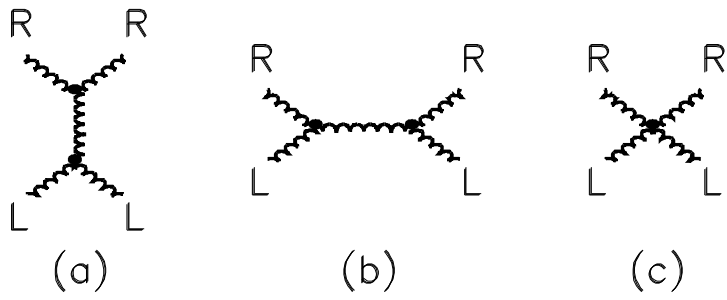
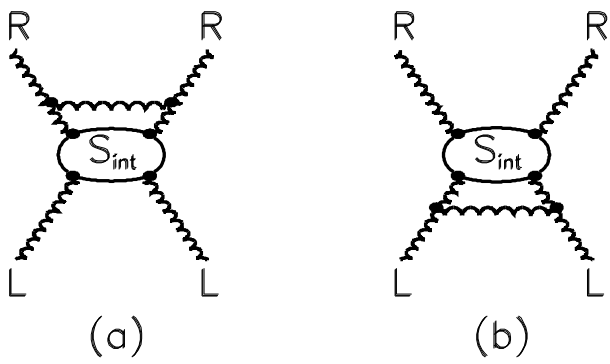
\end{document}